\documentclass[preprint]{aastex}
\usepackage{emulateapj5,apjfonts,psfig}

\newcommand{\etal}{{et~al.}}
\newcommand{\Msun}{M_\odot} 
\newcommand{\kms}{$\rm {km}~\rm s^{-1}$}

\slugcomment{{\it The Astrophysical Journal (Letters)}.}
\righthead{BLACK HOLE IN G1}
\lefthead{GEBHARDT, RICH, \& HO}

\begin{document}
 
\title{A 20 Thousand Solar Mass Black Hole in the Stellar Cluster 
G1\footnotemark[1]}

\footnotetext[1]{Based on observations made with the {\it Hubble Space
Telescope}, which is operated by AURA, Inc., under NASA contract
NAS5-26555.}
 
\author{Karl Gebhardt\altaffilmark{2}, R. M. Rich\altaffilmark{3}, and 
Luis C. Ho\altaffilmark{4}}

\altaffiltext{2}{Astronomy Department, University of Texas, Austin, TX
78723; gebhardt@astro.as.utexas.edu}

\altaffiltext{3}{UCLA, Physics and Astronomy Department, Math-Sciences 8979,
Los Angeles CA 90095-1562; rmr@astro.ucla.edu}

\altaffiltext{4}{The Observatories of the Carnegie Institution of
Washington, 813 Santa Barbara St., Pasadena, CA 91101; lho@ociw.edu}

\begin{abstract}
We present the detection of a $2.0(+1.4,-0.8)\times10^4~\Msun$ black
hole (BH) in the stellar cluster G1 (Mayall II), based on data taken
with the Space Telescope Imaging Spectrograph onboard the {\it Hubble
Space Telescope}. G1 is one of the most massive stellar clusters in
M31. The central velocity dispersion (25~\kms) and the measured BH
mass of G1 places it on a linear extrapolation of the correlation
between BH mass and bulge velocity dispersion established for nearby
galaxies. The detection of a BH in this low-mass stellar system
suggests that (1) the most likely candidates for seed massive BHs come
from stellar clusters, (2) there is a direct link between massive
stellar clusters and normal galaxies, and (3) the formation process of
both bulges and massive clusters is similar due to their concordance
in the $M_{\bullet}-\sigma$ relation. Globular clusters in our Galaxy
should be searched for central BHs.
\end{abstract}


\section{Introduction}

The questions of how the nuclei of galaxies form and why they contain
massive black holes (BHs) remain unsolved. However, the recent
discovery of a tight correlation between central black hole (BH) mass
and bulge velocity dispersion (hereafter the $M_{\bullet}-\sigma$
relation; Gebhardt et al. 2000b; Ferrarese \& Merritt 2000) does shed
some light on the evolutionary history of massive BHs and their host
galaxies.  Many theories (e.g., Silk \& Rees 1998; Haehnelt \&
Kauffmann 2000; Ostriker 2000; Adams, Graff, \& Richstone 2001)
predict such a correlation, and the exact details (i.e., slope and
normalization) can discriminate among the various models. Presently,
however, the data are inadequate to do this.  One difficulty is that
the galaxies studied so far have limited coverage in parameter space.
There are not enough observations at the low-dispersion end, and yet
this region provides the tightest constraints on determining both the
slope and offset. The main reason for this lack of low-dispersion
systems is that there are few that the {\it Hubble Space Telescope
(HST)}\ can reasonably observe.

Due to their high central densities and proximity, globular clusters
provide an alternative to studying galaxies to explore the low mass
end. Furthermore, there is evidence that at least some globular
clusters may be nuclei of accreted galaxies (Freeman 1993,
Ferguson~\etal\ 2002), and thus may contain central black holes if all
galaxies contain them (Magorrian~\etal\ 1998). The clusters G1 (Mayall
II) in M31 and M15 are excellent objects for these studies. They both
have high central densities and short central relaxation times ($10^7$
years for M15 and $10^8$ for G1).  Gebhardt~\etal\ (2000c),
van~der~Marel~\etal\ (2002) and Gerssen~\etal\ (2002) present results
for M15 showing that it likely contains a central BH of a few thousand
solar masses. If large (nonstellar) BHs exist in these stellar
clusters, which have low escape velocities and apparently lack dark
halos, they will pose a severe challenge to nearly every theory for
the formation of massive BHs.

The cluster G1 lies 40 kpc from the nucleus of M31, projected
approximately on its major axis.  It is the most luminous stellar
cluster in the Local Group and has a higher central surface brightness
than any Galactic globular cluster. Djorgovski et al. (1997) report a
velocity dispersion of 25 km s$^{-1}$ from ground-based spectroscopy,
and Meylan~\etal\ (2001) derive a total mass of $(7-17) \times
10^6M_\odot$, with the uncertainty due to their lacking a velocity
dispersion profile. To place G1 among Galactic globular clusters, we
note that the compilation of Trager, Djorgovski, \& King (1995) gives
no cluster with central surface brightness $<14.5$ $V$ mag
arcsec$^{-2}$, fully one magnitude fainter than G1 (Rich et al. 1996).
NGC 5139 ($\omega$~Cen) is about as massive and luminous as G1, but
its central surface brightness is 16.8 $\rm mag\ arcsec^{-2}$.  The
highest measured velocity dispersion for any Galactic globular cluster
is 18 km s$^{-1}$, for NGC 6441 and NGC 6388 (Pryor \& Meylan
1993). It is interesting to note that both G1 (Meylan~\etal\ 2001;
Ferguson~\etal\ 2002) and $\omega$~Cen (Freeman 1993) are possibly
nuclei of accreted galaxies.

The high luminosity and remarkable central surface brightness of G1
led us to propose to obtain STIS spectroscopy of its nucleus (GO-9099;
PI: Rich). This {\it Letter} reports the discovery of a
$2.0\times10^4\Msun$ BH in G1. Analysis of central population
gradients and detailed ground-based spectra will be reported in a
future paper (Rich et al. 2003).

\section{Data}

\subsection{STIS Observations}

We observed G1 using the Space Telescope Imaging Spectrograph (STIS)
with the G750M grating and the 0.1\arcsec$\times$52\arcsec\ slit on
2001 November 1 UT.  The position angle of the slit was 95\degr; since
the major axis of the cluster is at 120\degr\ (Meylan~\etal\ 2001),
the slit ran along an angle 25\degr\ up from the major axis. The
spectra cover 8276--8843 \AA\ with 0.554 \AA\ (19~\kms) per pixel and
a resolution of FWHM = 1.06 \AA\ or 37~\kms.  The total integration
time was 7.06 hr, divided into 20 exposures over two visits. We
dithered along the slit to aid in the removal of cosmic rays and hot
pixels. The dithering ranged between $\pm$1\arcsec, with non-integer
steps. These large dithers allow us to make a hot pixel map directly
from the data; this step is important because the hot pixels change
during every orbit, and, to ensure the best quality map, it is ideal
to use a hot-pixel map made from the data. The procedure for making
this map involves iterations whereby we make an initial
two-dimensional galaxy image that we subtract from the individual
images, which then are used to make a new hot-pixel map. Five
iterations are adequate to produce an accurate map. Pinkney~\etal\
(2002) describe this procedure in more detail.

We use contemporaneous flat images taken during each orbit to correct
for the pixel-to-pixel gain variation and fringing at these red
wavelengths.  The flat images are stable over many months and provide
flat-field correction to better than a percent.

\subsection{Kinematics}

We are able to obtain kinematics out to $\pm$1\farcs1. The modeling
described below utilizes the velocity profile directly. However, for
comparison with other work, Figure~1 plots the first two moments of a
Gauss-Hermite polynomial expansion. The estimate of the velocity
profiles relies on a non-parametric penalized maximum-likelihood
technique, which is described in Gebhardt~etal\ (2000a) and
Pinkney~\etal\ (2002). For the modeling, we use a symmetrized version
of the kinematics (solid lines in Fig.~1), which increases the
signal-to-noise for the estimate of the velocity profile.  Sampling
noise in the dispersion estimate---which can hamper measurements in
Galactic clusters (Dubath~\etal\ 1997)---is not significant for G1.
The central STIS pixel (0\farcs05$\times$0\farcs1) contains about
$10^4\Msun$ in stars projected into it, or 3000 $L_\odot$. If only
giant stars contributed light, which provides the extreme situation,
the central pixel would contain 30--100 stars, which is adequate to
overcome sampling noise.


\vskip 5pt \psfig{file=f1.eps,width=8cm,angle=0}
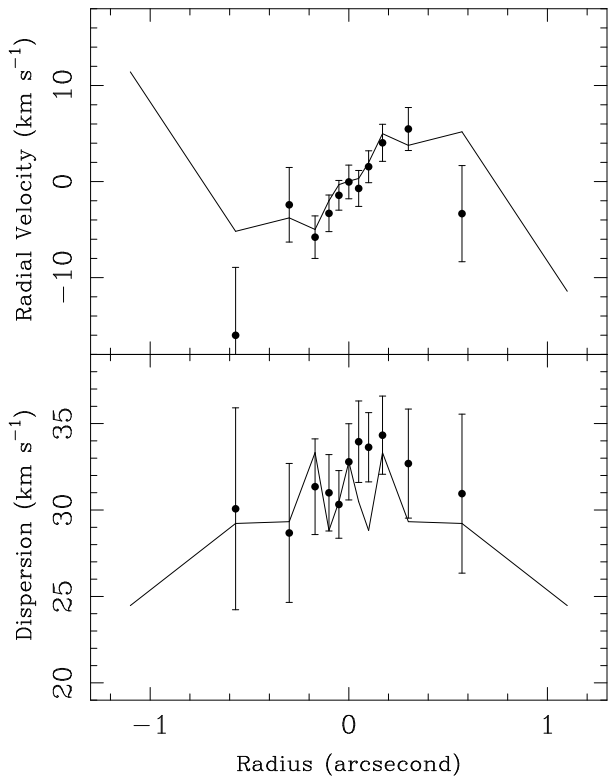
\figcaption{Radial velocity and velocity dispersion
profile of G1 from the STIS observations. The points and the
uncertainties come from the unsymmetrized measurements. The solid
lines come from the symmetrized estimate that are used in the
dynamical modeling. There is a single ground-based estimate of the
dispersion that is not shown here; Djorgovski~\etal\ (1997) measure
$\sigma = 25.1\pm1.7$~\kms\ in a 1\farcs2$\times$3\farcs0 aperture.
\label{fig1}}


\subsection{Photometry}

We use the images as described in Rich~\etal\ (1996). The surface
brightness profile at large radii ($r>$0\farcs3) is well measured
using either the data from Rich~\etal\ or Meylan~\etal\
(2001). However, in the central regions of G1, many of the exposure
where saturated, and we have to rely on the shorter exposures from
Rich~\etal\ (1996). These exposures consist of two 40~s exposures in
both the F555W and F814W filter. In both of these, the center is well
below the saturation limit. Unfortunately, these images where not
dithered and so we cannot reconstruct higher spatially sampled data
directly. We do not use deconvolved images for this
analysis. Deconvolution techniques accurately recover the intrinsic
surface brightness profile (Lauer~\etal\ 1998), but the lumpiness of
the G1 image, due to bright giant stars, may add significant noise to
the deconvolution. Such artifacts can be understood using simulations,
but for this initial analysis we rely on the observed images only
using the profile from Meylan~\etal\ Since the deconvolution will tend
to make the observed profile steeper, models using a deconvolved
profile will likely not change the best-fit BH mass but will make the
uncertainties smaller. Thus, we conservatively use the observed
profile.

We deproject the surface brightness profile to obtain the luminosity
density using a direct inversion of the Abel integral (Gebhardt~\etal\
1996). We assume a minor-to-major axis ratio of 0.75 that is constant
as a function of radius. This spheroidal distribution adequately
represents the configuration of G1 as measured by Meylan~\etal\
(2001). Changes in the assumed flattening have little effect on the
results below.

\section{Models}

The models are similar to those presented in Gebhardt~\etal\
(2002). They are axisymmetric, orbit-based models and so do not rely
on a specified form for the distribution function. Thus, for an
axisymmetric system, these models provide the most general
solution. The models require an input potential, in which we run a set
of stellar orbits covering the available phase space. We find a
non-negative set of orbital weights that best matches both the
photometry and kinematics to provide an overall $\chi^2$ fit. We vary
the central black hole mass and re-fit.

The orbit-based models store the kinematic and photometric results in
both spatial and velocity bins. For G1, we use 12 radial, 4 angular,
and 13 velocity bins. The data consist of the seven different STIS
positions along a position angle 25\degr\ up from the major axis and
one ground-based observation centered on the cluster. The point-spread
function for both {\it HST}\ and ground-based observations are
included directly into the models. The program matches the luminosity
density everywhere throughout the cluster to better than 0.5\%. The
quality of the fit is determined from the match to the velocity
profiles. The data points consist of $7\times13$ STIS velocity bins
plus the one ground-based dispersion, making 92 total points. However,
many of these points are correlated since the smoothing used for the
velocity profile extraction tends to correlate adjacent bins. The
reduction in the number of independent parameters is hard to estimate
but is generally around a factor of 2--4 (Gebhardt~\etal\ 2002).

Figure~2 plots a two-dimensional map of the different models and the
corresponding contours for $\chi^2$. The smallest value of $\chi^2$ is
17; given the 92 parameters, the reduction of the independent
parameters is about a factor of 5, higher than typical, which we
attribute to the small radial extent of the data. The two independent
parameters in the models are the BH mass and the stellar mass-to-light
ratio ($M/L$). The best-fit BH mass is
$2.0(+1.4,-0.8)\times10^4\,\Msun$ with $M/L_V$ = 2.6.
Figure~3 shows the one-dimensional plot of $\chi^2$ versus BH
mass. The difference in $\chi^2$ between the zero BH mass model and
the best fit is 3.0, implying a significance above 90\% for the BH
detection.


\vskip 5pt \psfig{file=f2.eps,width=8cm,angle=0}
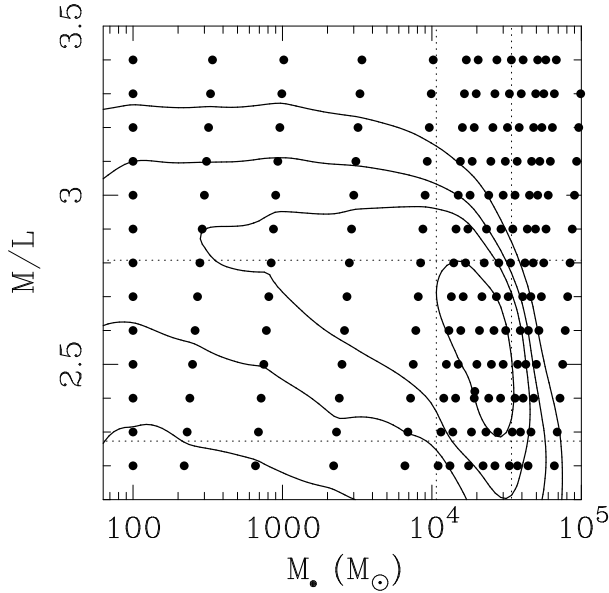
\figcaption{Two-dimensional plot of $\chi^2$ as a
function of BH mass and $M/L$ for G1. The points represent models that
we actually ran. The contours were determined by a two-dimensional
smoothing spline interpolated from these models, and represent
$\Delta\chi^2$ of 1.0, 2.71, 4.0, and 6.63 (68\%, 90\%, 95\%, 99\% in
projection). The vertical lines are the 68\% limit for the BH mass,
and the horizontal lines are the 68\% limit for $M/L$.
\label{fig2}}
\vskip 5pt



\vskip 5pt \psfig{file=f3.eps,width=8cm,angle=0}
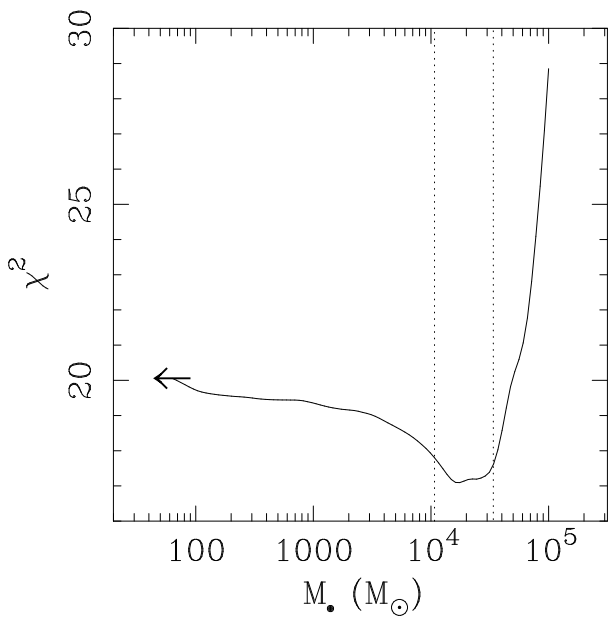
\figcaption{$\chi^2$ as a function of BH mass. We
have marginalized over $M/L$. The vertical dashed lines denote the
68\% confidence band quoted for the BH mass uncertainties. The arrow
on the leftmost point indicates that this point is actually at zero BH
mass, off the edge of the panel.
\label{fig3}}


\section{Results}

Our best-fit model has a BH mass of $2.0\times10^4\,\Msun$. We can
place this measurement on the $M_{\bullet}-\sigma$ relation
(Gebhardt~\etal\ 2000b; Ferrarese \& Merritt 2000) using the
ground-based measurement of $\sigma = 25.1\pm1.7$~\kms. Figure~4 plots
the $M_{\bullet}-\sigma$ correlation for nearby galaxies using the
compilation and the linear relation given in Tremaine~\etal\
(2002). G1 lies in excellent agreement with the extrapolation of the
linear fit to the local galaxies.



\psfig{file=f4.eps,width=8cm,angle=0}
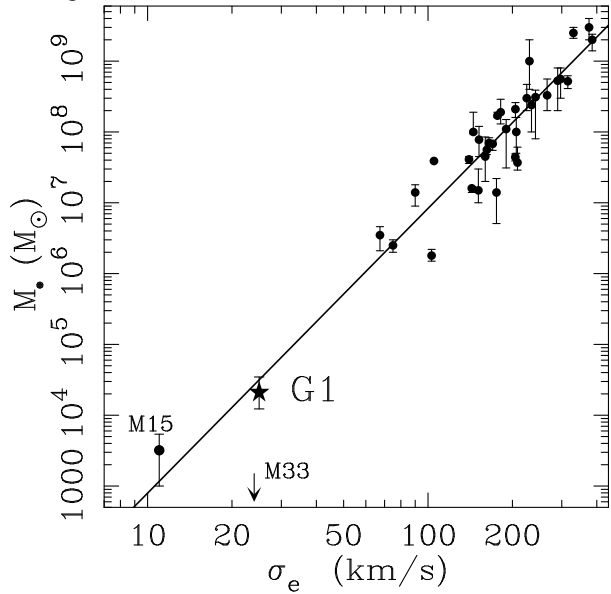
\figcaption{The $M_{\bullet}-\sigma$ correlation for
nearby galaxies, adapted from Tremaine~\etal\ (2002). We include here
the upper limit for M33 (Gebhardt~\etal\ 2001), the BH mass estimate
for the globular cluster M15 (van~der~Marel~\etal\ 2002 and
Gerssen~\etal\ 2002), and the mass for G1. The solid line is the
linear fit given by Tremaine~\etal, which does not include the cluster
G1 or M15.
\label{fig4}}


\section{Discussion and Conclusions}

G1 has traditionally been called a globular cluster. However,
Meylan~\etal\ (2001) offer the hypothesis that it is a nucleus of an
accreted small galaxy, similar to NGC~205. Furthermore,
Ferguson~\etal\ (2002) report the discovery of a disrupted galaxy near
G1, speculating that the two may have been part of the same system. A
similar origin for $\omega$~Cen has been proposed by Freeman (1993).
Evidence in favor of this interpretation for both clusters include
their large spread in metallicity and their exceptionally large
velocity dispersions relative to most globular clusters. If G1 is the
nucleus of an accreted galaxy, and if all bulge systems have BHs in
their centers (Magorrian~\etal\ 1998), then, to the extent that a
massive, bound cluster can be viewed as a ``mini-bulge,'' it is no
surprise that G1 has a BH as well. Combined with the results for M15,
it may be that {\it every}\ dense stellar system hosts a central BH.

There are significant consequences for these small systems having
central BHs.  First, it provides a direct link between stellar
clusters and galaxies. Galaxy correlation studies that include
globular clusters (Burstein~\etal\ 1997, Geha~\etal\ 2002) show that
they typically lie near to, but slightly offset from, the correlations
for the nearby galaxies. However, the large scatter prevents a
definitive comparison. The tightness of the $M_{\bullet}-\sigma$
relation allows us to explore their connection in better detail.  The
current results show that there is little difference between the
smallest and largest dense stellar systems.

Second, the existence of large (nonstellar) BHs in these small systems
constrains BH formation models. The low escape velocity of M15 and G1
($<$100~\kms) makes it difficult to grow a BH slowly over
time. Growing black holes from adiabatic accretion of gas is difficult
since globular clusters have a hard time holding onto any gas (namely
that from mass loss in evolved stars) due to their low escape
velocities. If BHs are grown from accretion of stars and stellar
remnants, the cluster must be able overcome the large recoil
velocities due to two-body interaction near the center.  A possible
solution to this dilemma is to have a large initial seed mass for the
BH that cannot subsequently get ejected from two-body interactions.
Miller \& Hamilton (2002) and Portegies~Zwart \& McMillan (2002)
discuss a mechanism in which massive BHs can exist in globular
clusters.

Third, one of the more important aspects of theories for the formation
of supermassive BHs in galaxies is that each has to start with a seed
BH. There are multiple explanations as to where these seeds come from;
whether they are primordial, created during the formation of the
galaxy, or formed in subsequent evolution is unknown. If small stellar
clusters contain BHs, due to their ages, they are a natural candidate
for the formation sites of seed BHs. A stellar cluster that formed
before the host protogalaxy collapsed could have easily donated its BH
to the galaxy center.  All of the theoretical models require only a
modest-sized BH to act as a seed, around $1\times10^4\,\Msun$ or less.

As an interesting counterexample, the galaxy M33 appears not to have a
BH.  Gebhardt~\etal\ (2001) measure an upper limit of $1500\Msun$. The
main difference between M33 and other galaxies with detected BHs is
that M33 does not have a a clear bulge component. However, M33 does
have a compact nucleus, whose stellar density is as high as those in
globular clusters. Thus, if any dense cluster has a BH then it is
puzzling that M33's nucleus has none. There are no obvious reasons for
this difference. However, M33's nucleus has a very different age than
G1; the former contains a significant population of stars younger than
a few Gyr (e.g., O'Connell 1983), whereas the latter is older than 10
Gyr (Meylan~\etal\ 2001). It is possible that either conditions to
make a massive BH were better in the past (i.e., different initial
mass function) or that M33's nucleus has not had enough time to create
one. In any event, we need more data on a larger set of nuclei and
clusters in order to explore this issue.

The model that we use for G1 assumes a constant stellar mass-to-light
ratio. The relaxation time for G1 near the center is short and we
expect heavy remnants there. By not including them, we overestimate
the BH mass. We estimate the effect by extrapolation of the
Fokker-Plank simulations of Dull~\etal\ (1997) for M15. They find that
$\sim 1000\,\Msun$ of remnants is in the central regions of that
cluster (see discussion in Gerssen~\etal\ 2002). G1 is $\sim$5 times
more massive than M15, implying that it has 5 times more remnants
given the same initial mass function. However, the relaxation times
are significantly longer in G1 (about a factor of ten), suggesting
that the presence of heavy remnants is even less of a
problem. Furthermore, Gerssen~\etal\ include models which incorporate
the appropriate mass-to-light variation and find that the required
black hole mass {\it increases} slightly. The reason is that even
though the remnants increase the mass-to-light ratio at the smallest
radius, the giant star cause it to drop towards the center since they
are centrally-concentrated. This drop in mass-to-light balances the
increase from the heavy remnants, thereby causing litle effect on the
BH mass.

A possible concern comes from comparing the {\it HST}/STIS dispersions
with that measured from the ground using different setups and
analysis. Therefore, we also ran models in which we use only the {\it
HST}\ data. We find essentially the same BH mass, but the confidence
band decreases slightly. The difference in $\chi^2$ between the no-BH
mass model and the best fit changes from 3.0 to 2.5. We find no reason
to suspect that either dispersion measurement is biased and use both
in the dynamical models.

The dynamical constraints for G1 can be improved with more extensive
ground-based kinematic observations (i.e., multiple position
angles). In addition to G1 and M15, there are a significant number of
globular clusters that can be exploited for these studies. Moreover,
ground-based observations should have sufficient spatial resolution to
measure a central BH. The most important challenge, however, lies in
understanding the contribution of heavy remnants. As explained above,
the impact of heavy remnants on G1 should not be severe because of its
long relaxation time, but more quantitative estimates of this effect
using evolutionary models would be highly desirable. Based on
experiments done so far in Gerssen~\etal\ (2002), it appears that the
BH mass estimates in M15 and G1 are unbiased.

\acknowledgements

K.~G. would like to thank Tod Lauer, Douglas Richstone, Christos
Siopis, John Kormendy and Paul Shapiro for many useful discussions. We
acknowledge grants under HST-GO-09099 awarded by the Space Telescope
Science Institute, which is operated by the Association of the
Universities for Research in Astronomy, Inc., for NASA under contract
NAS 5-26555.


\end{document}